\documentclass[sigconf,nonacm]{acmart}

\usepackage{amsmath,amsfonts}
\usepackage{algorithmic}
\usepackage{graphicx}
\usepackage{textcomp}
\usepackage{xcolor}
\usepackage{comment}
\usepackage[frozencache,cachedir=minted-cache]{minted}
\usepackage{pifont}
\usepackage{multicol}
\usepackage{hyperref}

\def\BibTeX{{\rm B\kern-.05em{\sc i\kern-.025em b}\kern-.08em
    T\kern-.1667em\lower.7ex\hbox{E}\kern-.125emX}}
\begin{document}

\newcommand{\code}[1]{\mintinline[breaklines, breakafter=_]{C++}{#1}}

\newcommand{\rr}{\textsuperscript{®}}
\newcommand{\tm}{\textsuperscript{™}}

\newcommand{\pgraphsp}{\vspace{0.33em}}
\newcommand{\paragrapht}[1]{\pgraphsp \indent \textbf{#1:}}

\newcommand{\cmark}{\ding{51}}%
\newcommand{\xmark}{\ding{55}}%



\ifx\nocomments\undefined
\newcommand{\nbc}[3]{
  {\colorbox{#3}{\bfseries\sffamily\scriptsize\textcolor{white}{#1}}}
  {\textcolor{#3}{\sf\small$\blacktriangleright$\textit{#2}$\blacktriangleleft$}}
}
\else
\newcommand{\nbc}[3]{}
\fi

\definecolor{babcolor}{rgb}{0.9,0.45,0.1}
\definecolor{wahrenscolor}{rgb}{0.06,0.07,0.62}
\definecolor{tdaviscolor}{rgb}{0.62,0.15,0.06}
\definecolor{jkcolor}{rgb}{0.11,0.5,0.06}
\definecolor{isaaccolor}{rgb}{0.5,0.1,0.6}
\definecolor{spencercolor}{rgb}{0.203,0.8203,0.9179} 

\newcommand{\bab}[1]{\nbc{BAB}{#1}{babcolor}}
\newcommand{\wa}[1]{\nbc{WA}{#1}{wahrenscolor}}
\newcommand{\willow}[1]{\nbc{WA}{#1}{wahrenscolor}}
\newcommand{\tad}[1]{\nbc{TAD}{#1}{tdaviscolor}}
\newcommand{\jk}[1]{\nbc{JK}{#1}{jkcolor}}
\newcommand{\iv}[1]{\nbc{TM}{#1}{isaaccolor}}
\newcommand{\spencerp}[1]{\nbc{SP}{#1}{spencercolor}}

\title{Binsparse: A Specification for Cross-Platform Storage of Sparse Matrices and Tensors}

\author{Benjamin Brock}
\affiliation{
  \institution{Intel Corporation}
  \city{San Francisco}
  \state{CA}
  \country{USA}
}
\email{benjamin.brock@intel.com}

\author{Willow Ahrens}
\affiliation{
  \institution{Massachusetts Institute of Technology}
  \city{Cambridge}
  \state{MA}
  \country{USA}
}
\email{willow@csail.mit.edu}

\author{Hameer Abbasi}
\affiliation{
  \institution{Quansight}
  \city{Beaufort}
  \country{Luxembourg}
}
\email{habbasi@quansight.com}

\author{Timothy A. Davis}
\affiliation{
  \institution{Texas A\&M University}
  \city{College Station}
  \state{TX}
  \country{USA}
}
\email{davis@tamu.edu}

\author{Juni Kim}
\affiliation{
  \institution{Massachusetts Institute of Technology}
  \city{Cambridge}
  \state{MA}
  \country{USA}
}
\email{junickim@mit.edu}

\author{James Kitchen}
\affiliation{
  \institution{Anaconda}
  \city{Austin}
  \state{TX}
  \country{USA}
}
\email{jkitchen@anaconda.com}

\author{Spencer Patty}
\affiliation{
  \institution{Intel Corporation}
  \city{Forest Grove}
  \state{OR}
  \country{USA}
}
\email{spencer.patty@intel.com}

\author{Isaac Virshup}
\affiliation{
  \institution{Chan Zuckerberg Initiative}
  \city{Redwood City}
  \state{CA}
  \country{USA}
}
\email{ivirshup@gmail.com}

\author{Erik Welch}
\affiliation{
  \institution{NVIDIA}
  \city{Austin}
  \state{TX}
  \country{USA}
}
\email{erik.n.welch@gmail.com}

\begin{abstract}
Sparse matrices and tensors are ubiquitous throughout multiple subfields of computing. The widespread usage of sparse data has inspired a multitude of in-memory and on-disk storage formats, but the only widely adopted storage specifications are the Matrix Market and FROSTT file formats, which are both ASCII text-based.
Due to the inefficiency of text storage, these files typically have larger file sizes and longer parsing times than binary storage formats, which directly store an in-memory representation to disk. This can be a major bottleneck; since sparse computation is often bandwidth-bound, the cost of loading or storing a matrix to disk often exceeds the cost of performing a sparse computation. While it is common practice for practitioners to develop their own, custom, non-portable binary formats for high-performance sparse matrix storage, there is currently no cross-platform binary sparse matrix storage format.  In this paper, we present Binsparse, a cross-platform binary sparse matrix and tensor format specification.  Binsparse is a modular, embeddable format, consisting of a JSON descriptor, which describes the matrix or tensor dimensions, type, and format, and a series of binary arrays, which can be stored in all modern binary containers, such as HDF5, Zarr, or NPZ.  We provide several reference implementations of Binsparse spanning 5 languages, 5 frameworks, and 4 binary containers. We evaluate our Binsparse format on every matrix in the SuiteSparse Matrix Collection and a selection of tensors from the FROSTT collection. The Binsparse HDF5 CSR format shows file size reductions of 2.4x on average without compression and 7.5x with compression.  We evaluate our parser's read/write performance against a state-of-the-art Matrix Market parser, demonstrating warm cache mean read speedups of 26.5x without compression and 2.6x with compression, and write speedups of 31x without compression and 1.4x with compression.
\end{abstract}

\maketitle



\begin{figure}[t]
  \centering
  \includegraphics[width=\columnwidth]{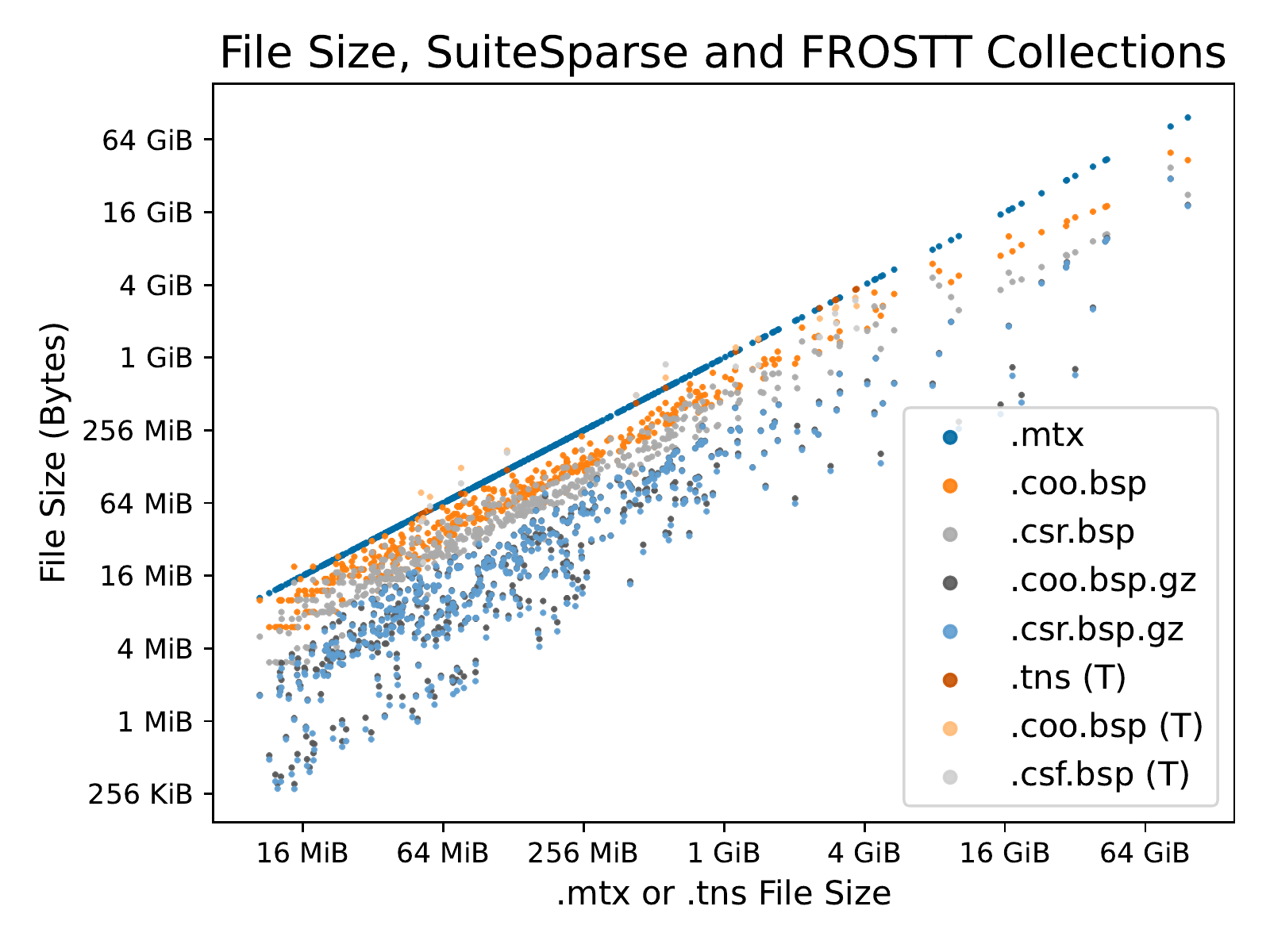}
  \caption{File sizes for all matrices in the SuiteSparse Matrix Collection with more than one million entries as well as 12 sparse tensors from the FROSTT Repository.  ``mtx'' indicates Matrix Market, while ``bsp'' indicates a Binsparse file format, ``csr'' indicates CSR format, ``coo'' indicates COO format, and ``gz'' indicates gzip compression is used. ``T'' indicates tensor data, ``.tns'' indicates the FROSTT format, while ``csf'' indicates the CSR-like compressed sparse fiber format.}
  \label{fig:file_size}
\end{figure}

\section{Introduction}
Sparse matrices and tensors are important data structures with many applications, including scientific simulation, data analysis, and machine learning. This diversity of applications has led to a diversity of sparse matrix and tensor frameworks, spanning multiple programming languages and architectures. Practitioners rely on file formats and specifications to transfer data between frameworks (e.g. Scipy\cite{virtanen_scipy_2020}, CuPy \cite{noauthor_cusparse_2024}, or PyTorch \cite{paszke_pytorch_2019}), languages (e.g. Matlab Tensor Toolbox \cite{bader_efficient_2008}, C++ TACO Sparse Tensor Compiler \cite{kjolstad2017taco}, or Python pydata/sparse tensors \cite{abbasi_sparse_2018}), and systems (e.g. from a local machine to a research cluster to cloud compute). Researchers use file formats to catalog reference datasets such as the SuiteSparse Matrix Collection~\cite{davis2011} or the FROSTT sparse tensor repository~\cite{frosttdataset}. Scientists checkpoint large simulations by storing simulation state (which may include sparse data) to disk. Machine learning experts distribute files containing neural network weights, which may be made sparse through pruning to save storage \cite{frankle_lottery_2019}.

A multitude of sparse matrix and tensor storage formats have been developed to support such wide-ranging use cases. \textbf{Practitioners must write bespoke code to target each format}. Each shape and nonzero sparsity pattern drives different dynamics in sparse operations, so specialized formats have been developed for each use case. In-memory formats (COO, CSR, CSF, DCSC, etc.) store sparse matrices during computation, but also serve as the inputs and outputs to library calls. Practitioners are required to read detailed documentation about small variations in memory layout or naming conventions, or write shim code to translate to an accepted format. On-disk file formats (such as Matrix Market~\cite{boisvert1996matrix} or Rutherford-Boeing~\cite{Duff:337273}) are used to write sparse matrices and tensors to the filesystem for storage, to be transferred to another system, or to be distributed widely. Each framework requires a separate parser for each file format. Practitioners often need to implement one-off parsers just to load their data.



In an effort to standardize, the community has converged on two on-disk file format specifications. Matrix Market~\cite{boisvert1996matrix} and FROSTT~\cite{frosttdataset} are the dominant sparse matrix and tensor file formats in use today. Unfortunately, both specifications utilize ASCII text for storage, which is inefficient and doesn't support direct serialization of common in-memory formats such as CSR.

While ASCII-based files have great portability---they can be used on any system that can parse text, regardless of the system's endianness or supported data types---they have fundamental inefficiencies. The first inefficiency is low information density: using 8-bit characters, which support 256 values, to store numeric digits, decimals, negative signs, and spaces wastes a large amount of the available character space.  As a result, ASCII-based files require about 1.7x\footnote{This is the average size reduction achieved by our Binsparse uncompressed COO format for all SuiteSparse Matrix Collection matrices with more than a million entries.  See Figure~\ref{fig:file_size} and Table~\ref{table:file_size}.}
more space than a binary coordinate-based format.  The second inefficiency is high computation burden: parsing text is computationally expensive.  When reading ASCII text, a parser must read in multiple characters, parse those into tokens, and then convert those tokens into numeric data types such as integers and floating point numbers. The third inefficiency is that ASCII-based files do not support direct serialization of in-memory formats. In fact, MatrixMarket and FROSTT only support sparse storage in coordinate format, typically sorted by convention in column order~\cite{davis2011,Kolodziej2019}, although this is not a requirement of the Matrix Market specification~\cite{boisvert1996matrix}. If another format is required, such as Compressed Sparse Row (CSR) or Doubly-Compressed Sparse Column (DCSC), an expensive format conversion process is required, which often involves storing an intermediate copy of the matrix in memory that must be sorted before constructing the matrix in the desired format.

Storing sparse matrices and tensors to disk can be a significant bottleneck.  Due to the low computational intensity inherent in sparse operations, they are typically memory bandwidth bound, meaning that the cost of computation is dominated by the cost of reading or streaming the sparse matrix from memory.  An often-overlooked consequence of this is that file IO---reading and writing sparse data to disk---can constitute a significant fraction of total runtime, in some cases being orders of magnitude more expensive than commonly optimized kernels such as SpMV, SpMM and SpGEMM.  This should come as no surprise: the runtime of an SpMV closely corresponds to the cost of reading the sparse matrix from memory.  Reading the sparse matrix from disk, which is two to three orders of magnitude slower than DRAM even on modern SSDs, will dominate the total runtime unless the sparse kernel is repeated many times.

Binary file formats, on the other hand, which directly store in-memory representations of sparse matrices to disk, are much more efficient.  In-memory binary representations almost always have smaller memory footprints than ASCII representations, therefore using less disk bandwidth.  In addition, as long as the on-disk binary representation is compatible with the system's in-memory binary data types, data can be directly copied from disk into memory, or vice versa, without requiring any manipulation.  However, due to compatibility issues between systems, as well as the large number of sparse matrix formats and data types, ASCII-based formats remain dominant: Matrix Market and FROSTT are the dominant formats used for communicating and storing sparse matrix and tensor data.  In practice, it is common for practitioners to develop their own hand-rolled binary sparse matrix or tensor formats, which can achieve much higher performance than even the fastest Matrix Market parsers~\cite{azad2022combinatorial,brock2024,wolfsonpou2019,10.1145/3293883.3295701,bharadwaj2023,stackoverflow2015,brock2021graphblas2}. However, due to incompatibilities between systems as well as the bespoke nature of these formats, they generally cannot be used in between libraries without manual intervention by the user to inspect and write a parser for the files.

In this work, we present Binsparse, a specification for cross-platform storage of sparse matrices and tensors in binary containers.  Binsparse is the result of a standardization effort to develop a specification~\cite{binsparse_spec_2024} and reference parsers.  The Binsparse specification is embeddable, meaning that it can be embedded inside pre-existing binary container libraries such as HDF5~\cite{Koranne2011,hdf5repo}, Zarr~\cite{zarr2023}, or even in-memory structures like DLPack~\cite{dlpack2025} for transferring in-memory sparse data between libraries.  It consists of a human-readable JSON file descriptor (see Listing~\ref{listing:basic_json}), which provides basic metadata about the stored matrix, such as its dimensions, number of entries, format, and structure.  Based on the JSON descriptor, there are defined to be one or more binary arrays stored in the file.  The Binsparse specification supports a large number of traditional in-memory sparse matrix formats as well as a hierarchical, declarative syntax for custom tensor formats.  This has the advantage of avoiding format conversion overhead, allowing libraries to read and write their desired sparse matrix formats directly to and from disk without the overhead of format conversion. Our contributions are as follows:

\begin{itemize}
  \item We present Binsparse, an embeddable, cross-platform specification for sparse matrix and tensor storage. Binsparse is the first such specification; there is no existing cross-platform standard for native binary storage of common sparse matrix formats. Note that our contribution is not the sparse data structures themselves, but rather the specification for storing them to disk in their native representations.
  \item We implement 7 different binsparse parsers, across 4 separate programming languages and 4 different binary containers.
  \item We convert all of the matrices in the SuiteSparse Matrix Collection~\cite{davis2011,Kolodziej2019} to Binsparse file formats based on COO and CSR.  We perform an analysis of all matrices with more than a million entries, finding that files are on average about 1.7x and 2.4x smaller without compression for COO and CSR, respectively, and 7.2x and 7.4x smaller with native HDF5 compression for COO and CSR, respectively.
  \item We convert a selection of tensors from the FROSTT Repository to Binsparse, observing that Binsparse CSF files are 6.7x and 1.2x smaller with and without compression, respectively.
  \item We analyze the read and write parse times of our reference HDF5 Binsparse parser compared to the fastest known Matrix Market parser~\cite{Lugowski_fast_matrix_market_Fast_and_2023}, finding that for our best performing format our parser achieves single-threaded average read speedups of 4.3x in cold cache experiments and 26.5x in warm cache experiments.  Average write speedups for the same format are 31x for warm cache experiments and 8.5x for cold cache experiments.
  \item We implement a parallel Binsparse reader, resulting in significant performance improvements for large files, particularly those using compression, with parallel reads on average about 2x faster than sequential reads.
\end{itemize}

\newcommand{\rotate}[1]{\rotatebox{90}{#1}}
\begin{table}[h]
\centering
\footnotesize
\caption{Currently Supported Binsparse Parsers}
\begin{tabular}{|l|l|c|c|c|c|c|c|c|}
\hline
\textbf{Parser Framework} & \textbf{Language} & \multicolumn{4}{c|}{\textbf{Containers}} & \multicolumn{2}{c|}{\textbf{Features}} \\ \cline{3-8}
        &                   & \rotate{\textbf{HDF5}} & \rotate{\textbf{Zarr}} & \rotate{\textbf{NPZ}} & \rotate{\textbf{In-Memory}} & \rotate{\textbf{Matrix}} & \rotate{\textbf{Tensor}} \\ \hline
SciPy                     & Python            & \cmark                 & \cmark                 &                        & \cmark                     & \cmark                  &                          \\ \hline
Finch.jl                  & Julia/Python             & \cmark                 &                        & \cmark                 & \cmark                     & \cmark                  & \cmark                   \\ \hline
Reference                 & C                 & \cmark                 &                        &                        &                             & \cmark                  &                          \\ \hline
Reference                 & C++               & \cmark                 &                        &                        &                             & \cmark                  &                          \\ \hline
pydata/sparse             & Python            & \cmark                 &                        &                        & \cmark                     & \cmark                  & \cmark                   \\ \hline
cupy                      & Python            &                 &                        &                        & \cmark                     & \cmark                  &                    \\ \hline
TACO                      & C++/Python            & \cmark                 &                        &                        &                            & \cmark                  & \cmark                   \\ \hline
\end{tabular}
\label{table:parser_support}
\end{table}
\section{Background and Related Work}

The ASCII-based Matrix Market format~\cite{boisvert1996matrix} is the dominant file format used for storing sparse matrices today.  Matrix Market grew out of a series of sparse matrix formats borne out of sparse matrix collections, including the Harwell~\cite[p. 76]{harwell1992}, Harwell-Boeing~\cite{10.1145/62038.62043} and Rutherford-Boeing~\cite{Duff:337273} sparse matrix collections, as well as the Matrix Market online repository of sparse matrices~\cite{osti_465805}.  The ASCII-based formats used by these collections include different variants of column compressed formats, finite element formats, and coordinate formats, with the coordinate-based Matrix Market format ultimately prevailing.  Text-based formats offer great portability, since they can be used on any system capable of parsing ASCII text, regardless of its endianness or supported data types.  Multiple highly optimized parsers for Matrix Market have been developed, including fast\_matrix\_market~\cite{Lugowski_fast_matrix_market_Fast_and_2023}, which uses the fastest available libraries for text parsing, and CombBLAS's distributed Matrix Market parser~\cite{azad2022combinatorial}, which uses multiple processes in parallel to read a Matrix Market file from a distributed filesystem.  However, even for frameworks that take the time to implement and use a highly optimized Matrix Market parser, it is common practice to convert commonly used matrices into a custom binary format that can be read directly from disk into memory, resulting in much faster parsing.  For example, CombBLAS itself implements a custom binary sparse matrix format~\cite{azad2022combinatorial} in addition to Matrix Market, as do many other distributed~\cite{brock2024,wolfsonpou2019} and single-node~\cite{10.1145/3293883.3295701,brock2021graphblas2} libraries, with users commonly hand-rolling their own custom binary formats as a matter of practice by either directly writing the binary arrays to disk or storing them inside a binary storage container such as HDF5~\cite{bharadwaj2023,stackoverflow2015}.  The majority of these formats involve writing the binary data structure directly from memory onto disk, which makes them extremely fast and even allows them to utilize memory-mapped files for efficient parallel and distributed reading.  However, the custom nature of these formats, combined with the fact that binary data dumped to disk is likely to encounter endianness and other compatibility issues when moved from system to system, means that these formats are non-portable.  Generally, when moving to a new system, files must be read from Matrix Market and new custom binary files must be generated if desired.  While using a cross-platform binary storage container such as HDF5~\cite{Koranne2011,hdf5repo} or Zarr~\cite{zarr2023} does provide portability for dense binary arrays, they do not explicitly support any major sparse formats.  Practitioners must instead manually examine such a container and then write code to extract the correct arrays, which has resulted in text-based formats overwhelming use whenever sparse matrices and tensors must be used portably.

For sparse tensors, the FROSTT~\cite{frosttdataset} sparse tensor repository's \code{.tns} format is the most commonly used format.  Like Matrix Market, FROSTT format is a text-based, whitespace separated list of nonzero indices and values in coordinate format, although typically with more than two dimensions. However, FROSTT does not specify the numeric type of values explicitly, nor does it specify the dimensions or number of nonzeros of the tensor.


For dense single and multi-dimensional arrays, a large variety of portable binary file formats are available.  Binary containers like HDF5~\cite{Koranne2011,hdf5repo}, NPY/NPZ~\cite{harris2020array} and Zarr~\cite{zarr2023} provide file formats that can hold multiple binary arrays inside a filesystem-like hierarchy of groups.  These containers' binary arrays can hold multi-dimensional dense tensors with a variety of layouts.  While they provide some of the core ingredients needed to store sparse tensors, they do not themselves support storing sparse matrices and tensors in a standard way.

Existing libraries like TileDB~\cite{papadopoulos2016tiledb} and (experimentally) HDF5~\cite{mainzer2019sparse} do offer chunk-based storage of sparse data.  Rather than natively storing traditional binary formats, chunk-based storage systems instead provide APIs for storing and performing computation over sparse data.  Users provide data as a set of sparse tuples, and these tuples are then stored using an opaque chunk-based data structure (in TileDB’s case, a tiled, compressed, columnar format; in HDF5’s case, a B-tree of compressed, dense, fixed-size chunks).  Unlike Binsparse, which preserves classical binary representations, chunk-based systems require a format conversion to read/write data in standard in-memory formats like CSR.  This makes them unsuitable for the large amount of applications that need to operate over sparse data in-memory using traditional sparse matrix formats.

\section{Binsparse Specification}
Binsparse is an embeddable specification for the storage of binary sparse matrix and tensor formats. A Binsparse file consists of two components: a JSON descriptor, which provides basic metadata about the matrix, such as its dimensions, number of stored values, structure, and format, and a collection of binary arrays, which are stored inside a binary container.  The JSON descriptor provides a convenient and human-readable description of the matrix stored and also provides a place (outside of the ``binsparse'' dictionary) to place other metadata about the matrix, such as SuiteSparse Matrix Collection~\cite{davis2011,Kolodziej2019} matrix metadata. JSON was chosen for its simplicity and wide adoption.

Being embeddable means that Binsparse can be used inside of pre-existing binary containers such as HDF5 and Zarr, and individual matrix components, such as the values or indices arrays, can be manually retrieved if desired.
Embeddability also provides for separation of concerns.  Since there already exist multiple well-supported libraries for storing binary data in a cross-platform way, handling endianness and other incompatibilities between systems, embeddability allows us to leverage existing work instead of defining yet another format.  Using third party binary container libraries also allows us to take advantage of any future performance or usability improvements that these widely used libraries receive.

\subsection{JSON Header}
Listing~\ref{listing:basic_json} provides an example of a Binsparse JSON header for the \code{Pajek/IMDB} matrix from the SuiteSparse Matrix Collection.  This header provides a number of keys with metadata about the matrix, including the matrix format, shape, number of stored values, and the data types of the binary arrays that compose the matrix.  Keys defined in the Binsparse Specification include:

\paragrapht{\code{format}} a string corresponding to one of the pre-defined traditional sparse matrix storage formats outlined in the Binsparse Specification~\cite{binsparse_spec_2024}.  The format determines what binary arrays will be provided and their meaning, as summarized in Table~\ref{table:formats}.

\paragrapht{\code{shape}} the number of rows and columns in the matrix.

\paragrapht{\code{number_of_stored_values}} the number of stored entries or ``nonzeros'' in the matrix.

\paragrapht{\code{data_types}} a JSON dictionary providing the types of the binary arrays specified by the format. The types are described with strings such as \code{int16} or \code{float32}. In order not to rely on the specifics of one binary container (such as HDF5), the binsparse spec itself defines each datatype, including signed and unsigned integers of various sizes as well as IEEE floats. Due to variations in how different programming languages handle boolean values, binsparse defines \code{bint8} to be an unsigned 8-bit integer, to be interpreted as a Boolean number however that is represented in the host language. Additionally, the datatype may include the modifiers \code{iso} and \code{complex}.  \code{iso} (for example \code{iso[bint8]}) indicates that all matrix entries hold the same value and that thus only a single value will be stored.  \code{complex} (for example \code{complex[float32]}) indicates that pairs of stored floating point numbers represent complex numbers.  These type modifiers allow Binsparse to efficiently store pattern and complex numbered matrices without requiring special support from the binary container library.

\paragrapht{\code{structure}} a string indicating the stored matrix values should be interpreted with a special structure, such as symmetric or Hermitian.

\paragrapht{\code{fill}} an optional Boolean value indicating whether the matrix has an implied fill value for unstored elements (for example $0$).  If present and \code{true}, the \code{fill_value} array must be stored in the binary container and contain the fill value.

\begin{table*}
  \centering
  \caption{Summary of specified binary arrays and their meaning for Binsparse pre-defined traditional sparse formats dense vector, dense matrix, compressed vector, coordinate, compressed sparse row, compressed sparse column, doubly compressed sparse row, and doubly compressed sparse column. Some supported column variants omitted for brevity.}
  \begin{tabular}{l | l l l l l l l l}
    & \multicolumn{8}{c}{Meaning by Format}\\
    \hline
    Array Name & DVEC & DMAT & CVEC & COO & CSR & CSC & DCSR & DCSC\\
    \hline
    \code{indices_0} & \xmark & \xmark & elem. index & row index & \xmark & \xmark & row indices & col. indices\\
    \code{indices_1} & \xmark & \xmark & \xmark & col. index & col. indices & row indices & col. indices & row indices\\
    \code{pointers_to_1} & \xmark & \xmark & \xmark & \xmark & row offsets & col. offsets & row offsets & col. offsets\\
    \code{values} & \cmark & \cmark & \cmark & \cmark & \cmark & \cmark & \cmark & \cmark\\
  \end{tabular}
  \label{table:formats}
\end{table*}

The JSON header defines one or more component binary arrays that make up the matrix data structure as detailed in Table~\ref{table:formats}. Binsparse supports pre-defined formats for the most common sparse matrix formats, as well as a hierarchical level-by-level custom format description for sparse tensors, described in Section~\ref{sec:custom_formats}. Binsparse uses level-based naming conventions for the binary arrays so that the names are consistent between predefined formats and their equivalent level-by-level analogues.  The pre-defined formats use a maximum of four arrays, \code{indices_0}, \code{indices_1}, \code{pointers_to_1}, and \code{values}. The array \code{indices_0} refers to the array that holds indices for the \emph{outer} (0th) level, which is row indices for row-based formats and column indices for column-based formats.  The array \code{indices_1} contains indices for the \emph{inner} (1th) level.  The \code{pointers_to_1} array is present when the inner (1th) level is compressed, and it holds offsets pointing into the arrays for that level (\code{values} and \code{indices_1} for pre-defined formats).  Pointers to the outermost (0th) level are never needed.

In our HDF5 reference parser described in Section~\ref{sec:hdf5_parser}, this JSON descriptor is stored as an attribute inside an HDF5 group (such as the root group), and its presence indicates that that group contains HDF5 datasets holding each of the component binary arrays.  To read a Binsparse matrix, a parser must parse the JSON descriptor, then read in the appropriate HDF5 datasets into memory.

\begin{listing}
\begin{minted}[fontsize=\footnotesize]{JSON}
{
  "binsparse": {
    "version":      "0.1",
    "format":       "COO",
    "shape":        [428440, 896308],
    "number_of_stored_values":      3782463,
    "data_types":   {
      "values":       "iso[bint8]",
      "indices_0":    "uint32",
      "indices_1":    "uint32"
    },
  },
  "comment":      "[omitted for brevity]"
}
\end{minted}
  \caption{The JSON header file for the Binsparse COO format of the SuiteSparse Matrix Collection's \code{Pajek/IMDB} matrix.}
\label{listing:basic_json}
\end{listing}

\subsection{Custom Sparse Tensor Formats}
\label{sec:custom_formats}

\begin{figure}
    \centering
    \includegraphics[width=\linewidth]{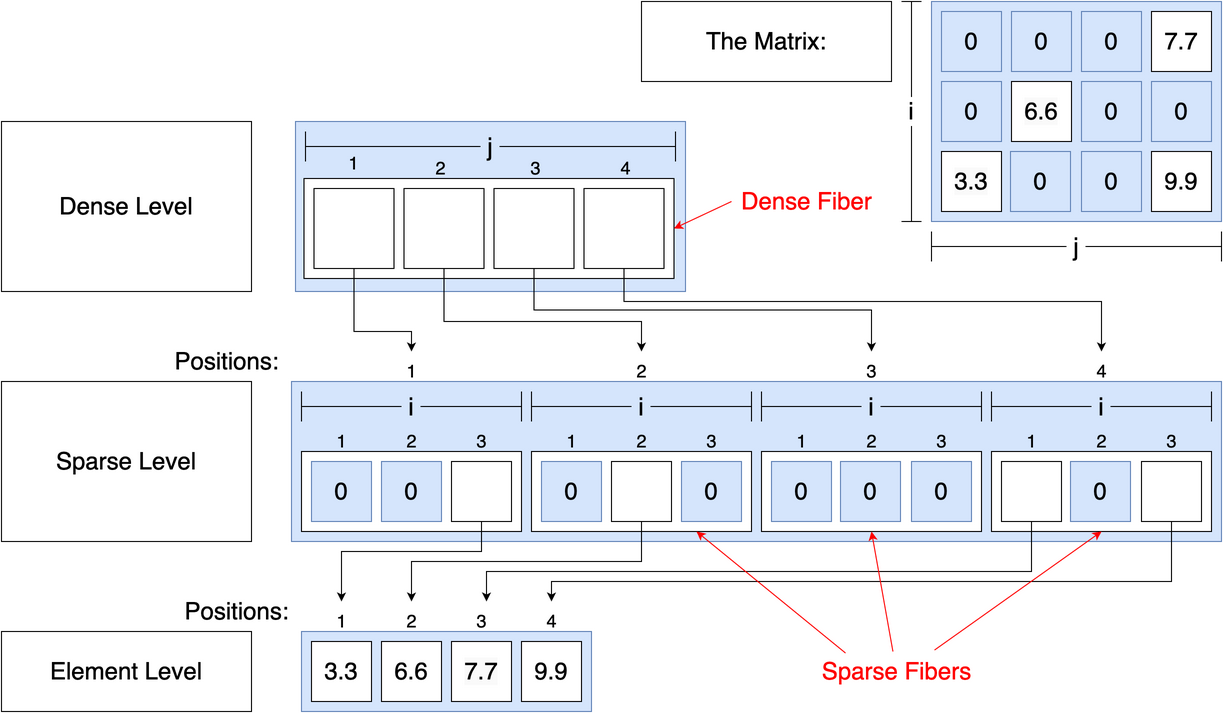}
    \caption{CSR matrix represented hierarchically}
    \label{fig:fibertree}
\end{figure}

Binsparse parsers can optionally support custom sparse tensor formats, which describe sparse tensors hierarchically. For this purpose, Binsparse adopts the \emph{fibertree} abstraction~\cite{sze_efficient_2020}, which is used in several popular tensor compilers including TACO~\cite{kjolstad2017taco}, MLIR sparse~\cite{bik_compiler_2022}, and Finch~\cite{ahrens2024finch}. We can understand these formats as arrays of arrays, where the parent array and
child arrays might use different formats. Each level of the tree corresponds to a dimension of the tensor. We refer to each array node in this tree as a \emph{fiber}. For example, Figure~\ref{fig:fibertree} shows how we might represent a CSR matrix as a dense
outer array which contains sparse inner arrays. To achieve efficient storage, all
arrays in the same level are stored contiguously in a specialized datastructure
called a level.

A level is a collection of zero or more fibers which all have the same format.
The elements of fibers in a level may be subfibers in a sublevel. The global
tensor we wish to store is represented by a level that holds a single root fiber.

Binsparse tensors are described under the \code{custom} tag, which stores a set of levels of type \code{dense}, \code{sparse}, or, for the innermost layer of scalar leaf values, \code{element}.  This set of level descriptors, along with a \code{transpose} key that describes the mapping from tensor dimensions to levels, defines a custom tensor format.  Listing~\ref{listing:custom_dcsc} shows a custom sparse tensor format equivalent to the pre-defined DCSC sparse matrix format.

Binsparse's custom sparse tensor formats mimic those used by sparse tensor compilers like TACO~\cite{kjolstad2017taco} and Finch~\cite{ahrens2024finch}, allowing them to write any in-memory format to disk.  This makes it straightforward to read or write in-memory tensors to disk by generating the equivalent Binsparse custom format. Whenever a custom format corresponds to a traditional sparse matrix format (such as CSR, COO, etc.), we substitute in the corresponding predefined alias.

We implemented our Binsparse parser for TACO using the reference HDF5 C implementation. Binsparse levels differ slightly from those of TACO~\cite{kjolstad2017taco} as Binsparse levels may span more than one dimension. While TACO uses a ``Singleton'' level to support COO formats (adding extra coordinates to an existing sparse level), Binsparse uses a 2-dimensional \code{sparse} level to support COO (the list of coordinates is contained in a single rank-2 sparse level). 

The custom level formats supported by Binsparse have a closer correspondence to the in-memory formats supported by Finch. Since Finch's custom formats use a column-major system, mapping indices to dimensions from right to left, the corresponding \code{transpose} key must be produced to properly map onto Binsparse's custom formats, which are row-major by default.  The default transpose key for Finch lists the dimensions in descending order.

\begin{listing}
\begin{minted}[fontsize=\footnotesize]{JSON}
"custom": {
  "transpose": [1, 0],
  "level": {
    "level_desc": "sparse",
    "rank": 1,
    "level": {
      "level_desc": "sparse",
      "rank": 1,
      "level": {
        "level_desc": "element",
      }
    }
  }
}
\end{minted}
  \caption{A custom sparse tensor format equivalent to the pre-defined DCSC.}
\label{listing:custom_dcsc}
\end{listing}

\section{Implementation}
In order to evaluate the Binsparse specification, we implemented parsers using several programming languages (C, C++, Julia, and Python) and binary container libraries (HDF5, Zarr, and NPZ) as well as several sparse matrix or tensor formats.  Given a file to parse, each of these parsers reads the corresponding matrix or tensor into an in-memory data structure.  Most of these programming languages use some form of \emph{runtime polymorphism}, meaning that the exact type of the matrix returned is not known until runtime.  This is due to the fact that a file may contain matrices of different formats and data types which will not be known until runtime, when the file is read.  The one exception is the C++ parser, which depends entirely on compile-time polymorphism, producing a matrix data structure whose types are known at compile time.  The C parser, which uses runtime polymorphism, leaving all these details until runtime, is also usable from C++.

For convenience, we will at times refer to a sparse matrix or tensor of an XYZ (COO, CSR, etc.) sparse format stored using an implementation of the Binsparse specification as a Binsparse XYZ matrix or tensor.

\subsection{HDF5}
\label{sec:hdf5_parser}

In HDF5, we use the convention of storing a Binsparse matrix inside an HDF5 group.  This could be the root group of the file, or it could be a manually created group, allowing multiple matrices to be stored in a single HDF5 file.  When a matrix is stored inside an HDF5 group, the group will contain a string attribute named \code{binsparse}.  The \code{binsparse} attribute will contain the JSON descriptor for the Binsparse file stored in that group.  Given an HDF5 group, the parser then only needs to parse the JSON descriptor, then read the component binary arrays stored as HDF5 datasets within the group.  The process for writing is similar, except that the string attribute and binary arrays must be written to the HDF5 group.  When writing binary arrays, we by convention write using the standard little endian HDF5 type corresponding to each data type.  HDF5 handles endianness automatically, as when reading in binary arrays, we simply have to tell HDF5 that we wish to read the array into a native C array of the corresponding data type.  HDF5 allows us to transparently apply native gzip compression to each dataset if desired, which we expose as an optional parameter.  In our reference parser, we pick a chunk size equal to 1 MiB, which is the default and generally recommended chunk size.

When reading or writing a file, our parsers accept a file name along with an optional group name for the HDF5 group of the desired Binsparse matrix.  In addition to reading and writing matrices, our parser can also iterate through all HDF5 groups in a file, looking for any Binsparse matrices.  We provide a \code{bsp-ls} program that lists all Binsparse matrices within an HDF5 file.

Using our HDF5 parser, we have converted every matrix in the SuiteSparse Matrix Collection into Binsparse COO and CSR formats using the HDF5 binary container.  Since Binsparse allows multiple matrices to be stored per file, this allows us to store all of the matrices found in each matrix tarball, including auxiliary matrices, inside a single HDF5 file.  We store the primary matrix in the root group, and each of the auxiliary matrices inside an HDF5 group with the auxiliary matrix's name.  We apply a ``declamping'' process to set each value greater than or equal to $1 \times 10^{308}$ equal to infinity, consistent with the official SuiteSparse Matrix Collection Matrix Market parser.  (Clamping was originally applied to store infinity values using the compact string \code{1e308}.)  We also apply a memory footprint minimization process to pick the optimal data types for storage as further described in Section~\ref{sec:minimization}.  We performed validation between the converted Binsparse files and original Matrix Market files, finding the corresponding in-memory representations to be bitwise identical.

\subsection{NPY}

The NPY format~\cite{npy_format} stores binary NumPy array data on disk together with a serialized Python dictionary of metadata.  The format is simple enough that there are multiple header-only NPY parsers in C, and, perhaps bolstered by the success and ubiquity of Python, the format has found its niche as a lightweight tensor format. The NPZ format simply stores multiple NPY files in a ZIP archive~\cite{npy_format}. NPZ allows users to store multiple dense arrays within the same file. We implemented a NPY Binsparse parser, which uses separate NPY files to store each of the component data arrays specified by the Binsparse format along with a raw JSON file containing the Binsparse descriptor.

\subsection{In-Memory Interchange}
Traditionally, when calling a library routine from one sparse framework using data from another, we must convert the objects or structs from the current framework to those of the framework we wish to call. This often requires specialized code for each situation. An in-memory interchange format is standard across all frameworks, so we can seamlessly copy data between two frameworks which may have been designed independently. 

Because the Binsparse specification provides a standard naming convention for the constituent data arrays underlying in-memory formats, it can be used for such in-memory exchange. Binsparse provides the means to standardize the names and descriptions of the tensor objects. This has been implemented and tested in multiple Python array libraries. We tested this in SciPy, CuPy, and pydata/sparse. The proposed protocol passes a Python dictionary containing a binsparse file descriptor and several arrays named with Binsparse conventions. Each array's data is stored using the zero-copy binary array interchange format DLPack~\cite{dlpack2025}, and the file descriptor is itself stored as a Python dictionary instead of a JSON file. This bypasses the need to write specialized code to move sparse arrays between Python sparse array implementations and allowing for in-memory zero-copy exchange of sparse arrays.

\section{Evaluation}

We evaluated an implementation of the Binsparse specification for sparse matrices by comparing our reference C parser against a state-of-the-art Matrix Market parser, fast\_matrix\_market, which to the best of our knowledge is the fastest Matrix Market parser available~\cite{Lugowski_fast_matrix_market_Fast_and_2023}.  First, we convert every matrix in the SuiteSparse Matrix Collection~\cite{davis2011,Kolodziej2019} from Matrix Market format into several Binsparse matrix formats using our reference parser and a variety of settings in order to examine the file sizes achieved.  Then, we evaluate the read and write parsing performance under warm cache scenarios, comparing the achieved performance to fast\_matrix\_market. We also evaluate an implementation of the Binsparse specification for sparse tensors by comparing our TACO parser against the state of the art FROSTT format parser splatt~\cite{frosttdataset,smith_splatt:_2015}.  We converted all tensors from the FROSTT dataset containing less than two billion nonzeros due to a limitation in TACO, which hard codes 32-bit signed integers for all indices.

Experiments were performed on a system equipped with a 6-core Intel\rr{} Core\tm{} i5-9600K CPU, 128~GB of DDR4 memory, and a 2~TB Samsung 970 EVO Plus NVMe\rr{} M.2 SSD.

\subsection{Matrix Conversion}
To convert matrices from Matrix Market representation to a Binsparse sparse format, we first read them into an in-memory COO matrix data structure that can then be directly written to disk as a Binsparse COO matrix or converted to another format and written to disk using a Binsparse parser. To read the file into memory, we must first pick explicit types for storing indices and values.  For indices, our parser always picks optimal integer widths based on the dimensions of the matrix.  For example, if both matrix dimensions are below $2^8$, \code{uint8} will be used, if they are larger than $2^8$ but below $2^{16}$, \code{uint16}, and so on.  For the value types, we can only observe the Matrix Market field type, which is one of \code{real}, \code{complex}, \code{integer}, or \code{pattern}.  When initially reading into memory, our parser reads in \code{real} values as \code{float64}, \code{complex} values as \code{complex[float64]}, \code{integer} values as \code{int64}, and \code{pattern} values as a single \code{bint8} value equal to \code{true}, with the matrix marked as ISO.  The structure of the matrices is carried over using the corresponding Binsparse \code{structure} value.

In order to store the matrices as efficiently as possible, when converting the SuiteSparse Matrix Collection we also apply a "downcasting" pass to the values array.  This pass iterates through every stored value, checking whether each value in the matrix is equivalent when cast to a smaller width (e.g. \code{float32} instead of \code{float64}).  If so, the entire values array is downcast to save space.  We then write the matrices to disk.  For each matrix, we evaluate four Binsparse formats: COO, COO with compression, CSR, and CSR with compression.  For compression we use HDF5's built-in gzip compression with the fastest level of compression (\code{-1}) selected.  We evaluated higher levels of compression in our testing, but found that higher levels of compression had significantly slower read/write times with minimal improvements in file size, so only the fastest level (\code{-1}) of compression is shown here.

\label{sec:minimization}
The file sizes for every matrix in the SuiteSparse Matrix Collection with more than one million stored values are shown in Figure~\ref{fig:file_size}.  To simplify the analysis, here we examine only the primary matrix for each SuiteSparse Matrix Collection tarball, along with a Binsparse conversion of that matrix.  As shown in the plot, the Binsparse files, which store raw binary data using HDF5, overwhelmingly have smaller file sizes.  As listed in Table~\ref{table:file_size}, average reduction in file size is 1.7x and 2.4x for COO and CSR formats without compression and 7.2x and 7.5x when compression is used.  Files are reduced in size up to 5.5x for CSR without compression, and up to 53.6x for CSR with compression.  Only four matrices had larger Binsparse files than the original Matrix Market files, all with the uncompressed COO format.  We examined these four matrices, and found they were all small matrices (less than 4 million entries) with the \code{real} field type.  These matrices all contain real values, for example $0.7$ and $0.8$, that have compact ASCII representations but are not precisely representable in \code{float32} (or indeed \code{float64}).  It should be noted that since the matrices in the Matrix Market files in the SuiteSparse Matrix Collection were generated with the intention of being used as archival material, their parser performs a minimization process that exhaustively tests format strings to identify the smallest printf width specifier that does not reduce accuracy.  This minimization process is costly, but ensures that the SuiteSparse Matrix Collection contains the smallest possible Matrix Market file for a given matrix.

\begin{table}
  \centering
  \caption{File size reductions of Binsparse file formats for SuiteSparse Matrix Collection Matrices with at least one million entries.}
  \begin{tabular}{l | r r r}
    File Format & Avg Reduction & Max Reduction & Min Reduction\\
    \hline
    .coo.bsp & 1.7x & 3.8x & 0.8x\\
    .coo.bsp.gz & 7.2x & 48.9x & 2.4x\\
    .csr.bsp & 2.4x & 5.5x & 1.1x\\
    .csr.bsp.gz & 7.5x & 53.6x & 2.3x\\
  \end{tabular}
  \label{table:file_size}
\end{table}

\begin{table}
  \centering
  \caption{File size reductions of Binsparse file formats for FROSTT Repository tensors with at least one million entries and less than two billion entries.}
  \begin{tabular}{l | r r r}
    File Format & Avg Reduction & Max Reduction & Min Reduction\\
    \hline
    .coo.bsp & 0.9x & 1.4x & 0.6x\\
    .coo.bsp.gz & 8.9x & 17x & 4.7x\\
    .csf.bsp & 1.2x & 2.5x & 0.6x\\
    .csf.bsp.gz & 6.7x & 15.3x & 3x\\
  \end{tabular}
  \label{table:file_size_tensor}
\end{table}

We recognize that the ad hoc use of binary file tools could achieve similar improvements over the Matrix Market file format and parser.  The benefit of the Binsparse specification is that we are proposing a specification that provides a common framework around using such tools to represent sparse matrices (and tensors).  This allows users across different platforms, languages and tool-chains to take advantage of such improvements in file size and parsing speed.

\subsection{Tensor Conversion}
The FROSTT format (\code{.tns}) does not specify the tensor's dimensions or datatype.  As is the convention in FROSTT parsers, when converting tensors to Binsparse format, we choose the tensor shape based on the largest coordinates in each dimension. Due to our parser's integration with TACO, we use 32-bit integers for all indices.  When selecting the value type, as is common practice, we attempt to parse the stored values as boolean, integer, float, or complex, picking the first successful parse as the datatype for the tensor.  We store tensors using the Binsparse COO and compressed sparse fiber (CSF) formats with and without compression.  Average, minimum, and maximum file size reductions are shown in Table~\ref{table:file_size_tensor}.  Uncompressed COO and CSF file sizes are shown in Figure~\ref{fig:file_size}.

\subsection{Matrix Parsing}
In addition to file size, we also evaluated the parsing performance of our reference Binsparse C parser, benchmarking both read and write performance.  For read experiments, we ran both warm and cold cache experiments.  For cold cache experiments, we flush the filesystem cache\footnote{The filesystem cache was flushed using the command \code{sync && sudo echo 3 > /proc/sys/vm/drop_caches}.} before every read to ensure that the file is read directly from disk.  In warm cache experiments, the file is read in once to warm the filesystem cache before any measurements are taken.  For write experiments, we perform both flushed and unflushed experiments.  For flushed experiments, we flush all outstanding writes to disk\footnote{We flushed all writes to disk using the command \code{sync}.} after the file has been written, including the flush in the measured time.  In unflushed experiments, we simply write the file, without forcing the writes to be flushed to disk.

Cold cache and flushed experiments ensure that our measurements actually measure the cost of reading and writing to disk, meaning that disk bandwidth can become a bottleneck.  This is arguably the most representative use case, since a large sparse matrix file is unlikely to be cached in memory unless it is being read frequently and memory utilization is low, providing enough free memory for the operating system to leave the file in cache.  However, warm cache and unflushed experiments are represented in some use cases, such as when a small matrix is read multiple times or writes are performed intermittently, allowing data to be written to disk in the background while other work is performed.  Warm cache and unflushed experiments also better measure the raw cost of the parser itself, since disk bandwidth does not present a bottleneck.

The experiments below were performed on the system described earlier in this section, which is equipped with a 2~TB Samsung 970 EVO Plus NVMe\rr{} M.2 SSD.  This SSD has manufacturer reported read and write bandwidths of 3.5~GB/s and 3.3~GB/s, respectively~\cite{samsung2024}.  The SSD was connected to a PCIe Gen3 x4 M.2 slot.  All Matrix Market experiments use fast\_matrix\_market, with reading and writing performed via the functions \code{read_matrix_market_triplet} and \code{write_matrix_market_triplet}.  In order to perform a fair comparison, we used the same (optimal) integer and floating point widths for indices and values when reading or writing Matrix Market matrices that we used with Binsparse matrices.  The option \code{generalize_symmetry} was set to \code{false} so that symmetric matrices would not duplicate triangular entries in memory.  For single-threaded experiments, multi-threading was disabled, while in our parallel read experiments it was enabled with the number of threads set to 6.  Each runtime reported is an average over ten trials.

\subsection{Read Experiments}
\begin{figure*}
  \centering
  \includegraphics[width=0.497\textwidth]{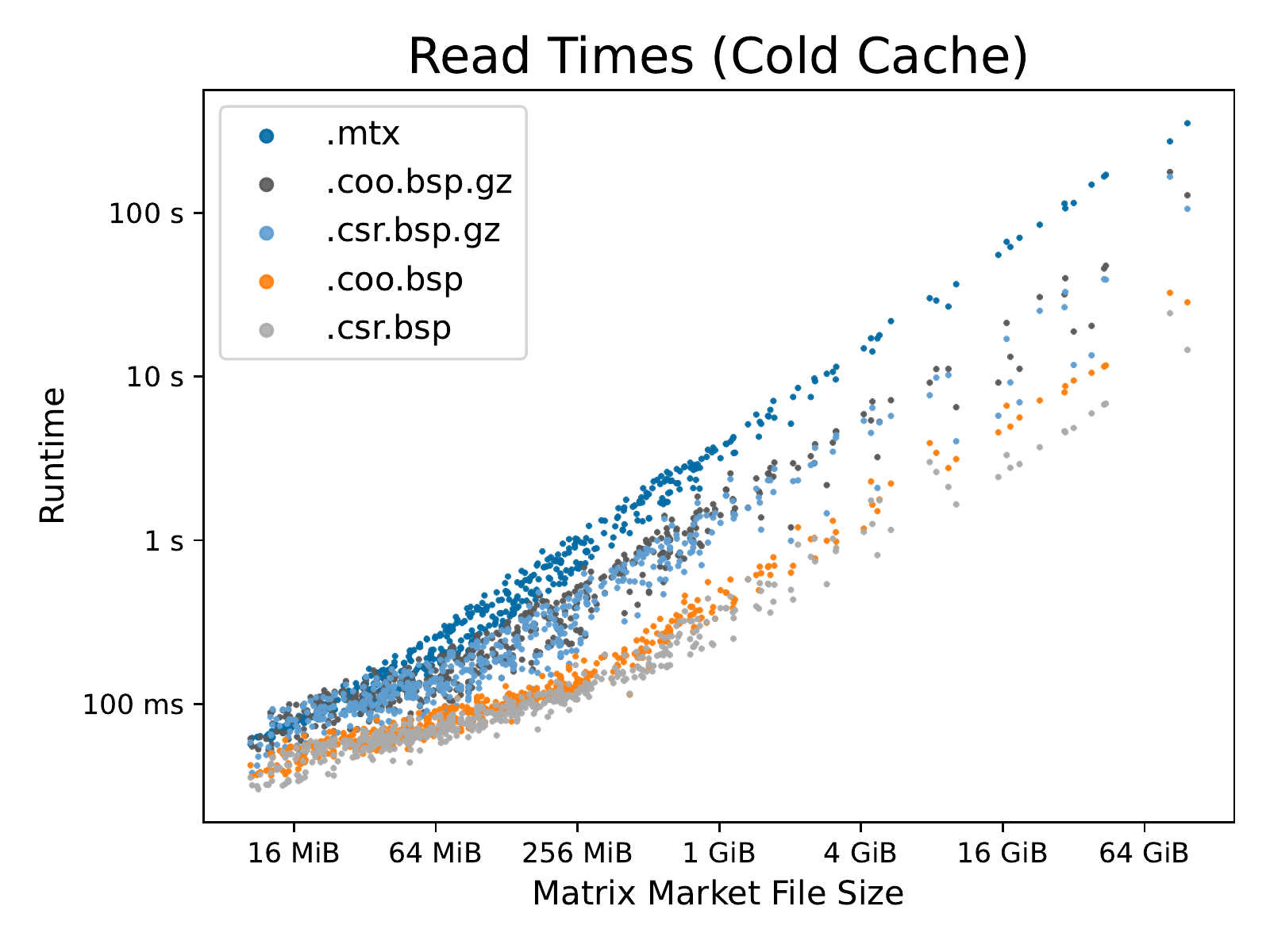}
  \includegraphics[width=0.497\textwidth]{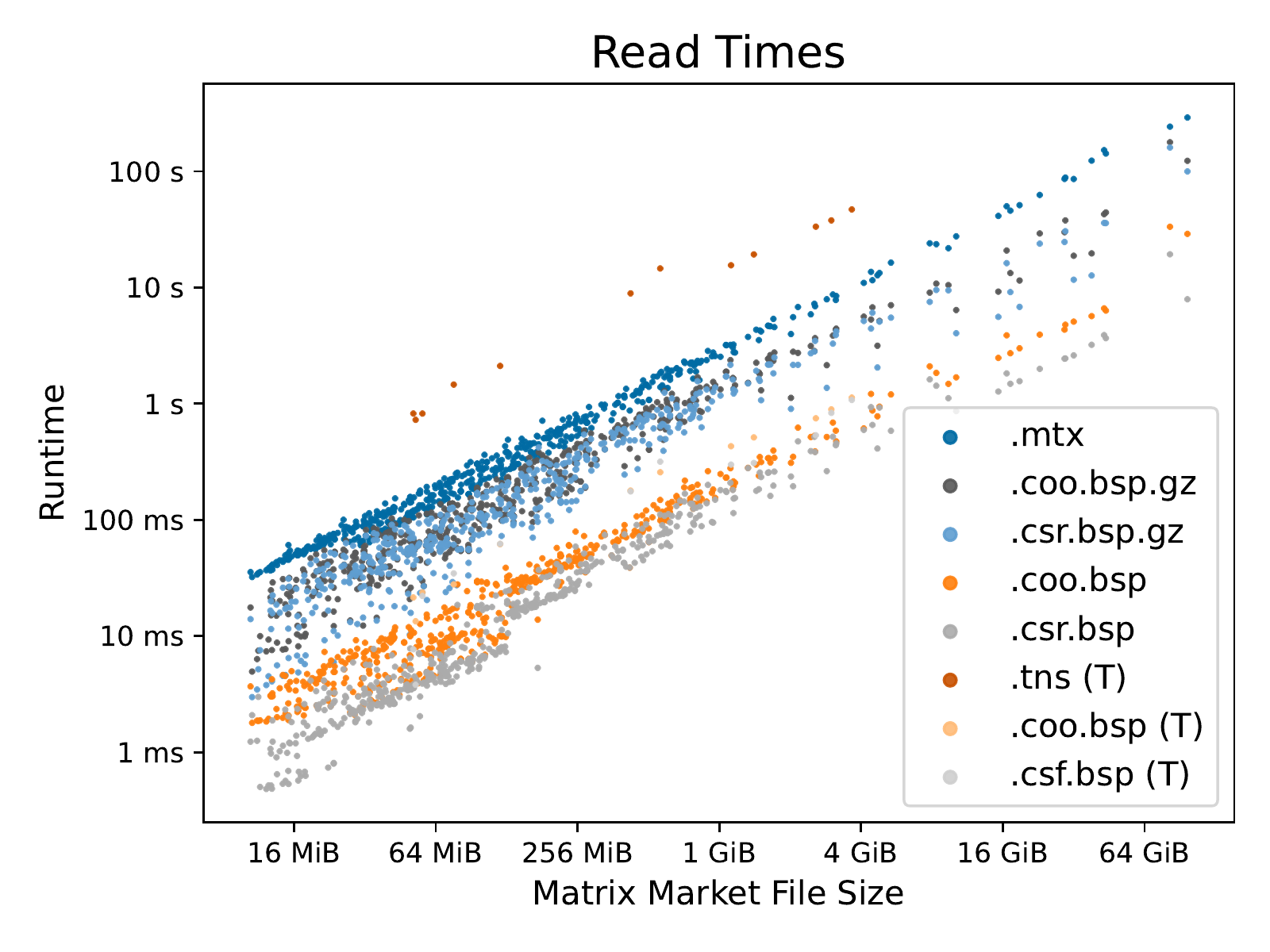}
  \caption{Single-threaded cold and warm cache read times for all matrices in the SuiteSparse Matrix Collection with more than one million entries.  ``mtx'' indicates Matrix Market, while ``bsp'' indicates a Binsparse file format, ``csr'' indicates the CSR format is used, ``coo'' indicates the COO format is used, and ``gz'' indicates gzip compression is used.}
  \label{fig:read_times}
\end{figure*}

Figure~\ref{fig:read_times} plots single-threaded time taken to read every sparse matrix in the SuiteSparse Matrix Collection with more than one million entries, including both cold cache and warm cache experiments.  Each dot represents the time taken to read a single matrix, averaged over ten runs, with time in the y-axis and the size of the matrix's Matrix Market file on the x-axis.  Average, minimum, and maximum speedups are shown in Table~\ref{table:read_speedups}.  Binsparse read times are broadly faster than Matrix Market read times.  Uncompressed Binsparse formats are always faster than Matrix Market, with the minimum speedups being 1.2x and 7.3x for COO uncompressed with cold and warm cache, respectively.  The highest average speedups are for CSR uncompressed, with 4.3x and 26.5x average speedup for cold and warm cache, respectively.  Binsparse formats achieve higher speedups in warm cache experiments, which stress the computational efficiency of the parsers, because they do not have to parse ASCII text, which makes Matrix Market parsing more computationally expensive.  Instead, they simply copy binary data from disk into memory using HDF5.  In cold cache experiments, where read bandwidth is lower, the gap narrows somewhat.

Reading compressed matrix formats is slower than reading uncompressed formats, since additional computation is required to decompress the data.  For very small matrices, reading compressed Binsparse files may be slower than Matrix Market files.  However, as matrix size increases, the gap widens, with Binsparse compressed CSR achieving up to 11x and 14.5x speedup for cold and warm cache, respectively, on larger matrices.

Table~\ref{table:read_speedups_tensor} shows the average speedups for Binsparse COO and CSF formats over FROSTT's splatt parser, which are 43x and 56x, respectively.  Average speedups for compressed formats are both above 9x.


\begin{figure*}
  \centering
  \includegraphics[width=0.497\textwidth]{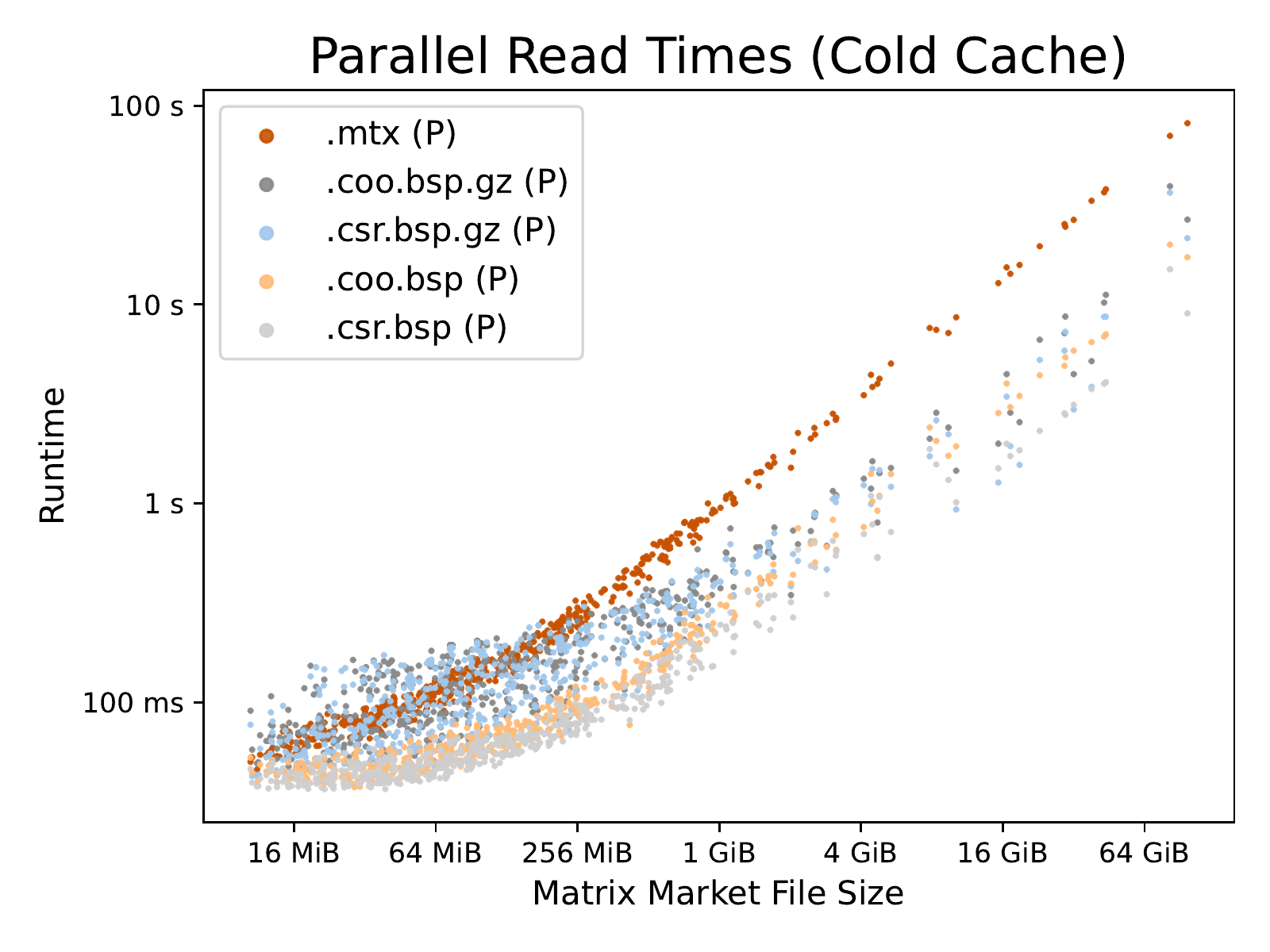}
  \includegraphics[width=0.497\textwidth]{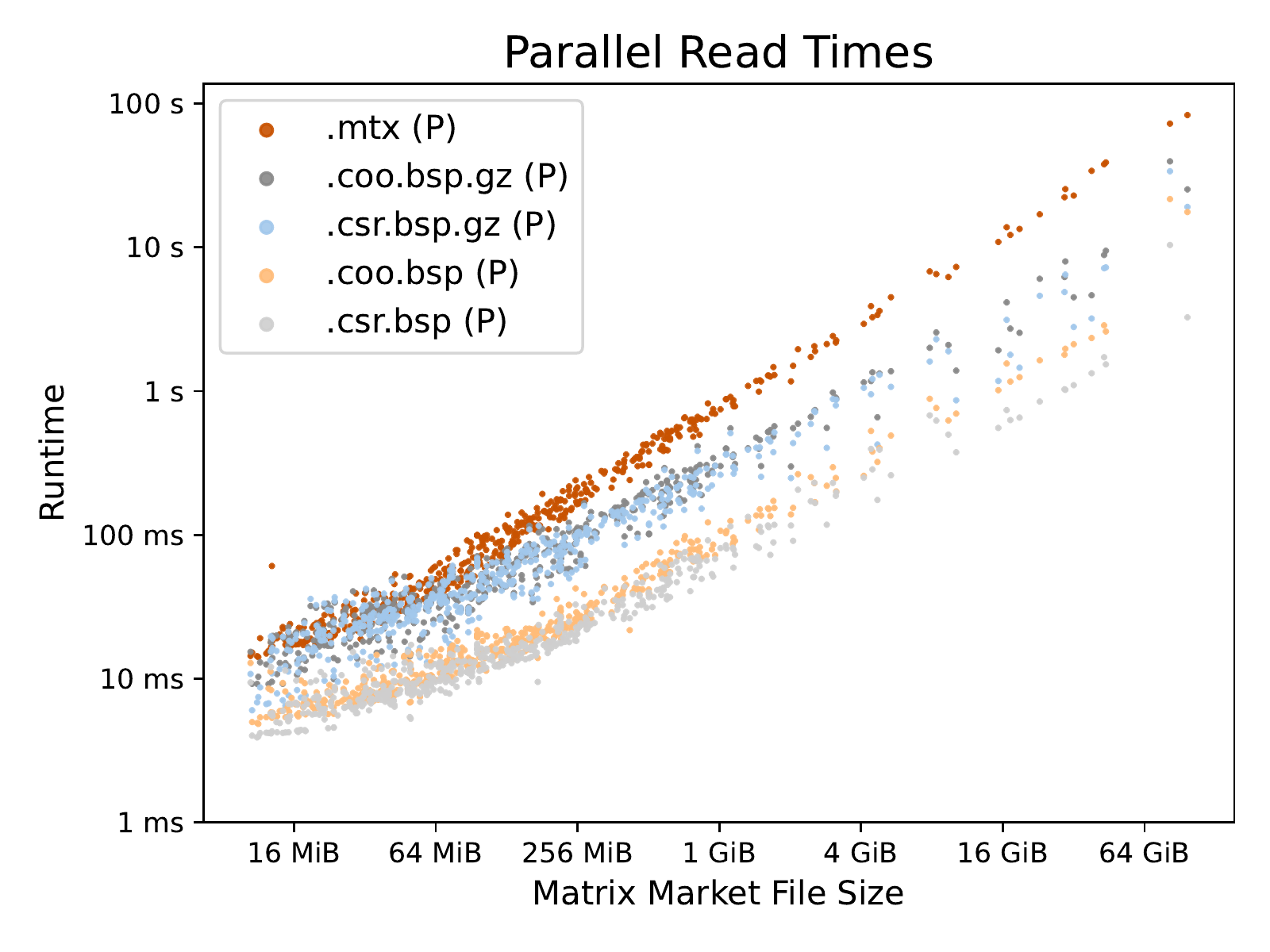}
  \caption{Cold and warm cache \emph{parallel} read times for all matrices in the SuiteSparse Matrix Collection with more than one million entries.  ``mtx'' indicates Matrix Market, while ``bsp'' indicates a Binsparse file format, ``csr'' indicates the CSR format is used, ``coo'' indicates the COO format is used, ``gz'' indicates gzip compression is used, and ``P'' indicates parallel.}
  \label{fig:read_times_mt}
\end{figure*}

\begin{table}
  \centering
  \caption{Single-threaded read time speedups over fast\_matrix\_market for cold cache and warm cache experiments on SuiteSparse Matrix Collection matrices with at least one million entries.  Cold cache speedups are shown first, with warm cache speedups following in parantheses.}
  \begin{tabular}{l | r r | r r | r r}
    File Format & \multicolumn{2}{c|}{Avg Speedup} & \multicolumn{2}{c|}{Max Speedup} & \multicolumn{2}{c}{Min Speedup}\\
    \hline
    .coo.bsp & 3.7x & (15.9x) & 14.6x & (37.7x) & 1.2x & (7.3x)\\
    .coo.bsp.gz & 1.7x & (2.2x) & 7.3x & (7.2x) & 0.7x & (1.1x)\\
    .csr.bsp & 4.3x &(26.5x) & 24.9x & (98.2x) & 1.3x & (11.6x)\\
    .csr.bsp.gz & 1.8x &(2.6x) & 11x & (14.5x) & 0.7x & (1.1x)\\
  \end{tabular}
  \label{table:read_speedups}
\end{table}

\begin{table}
  \centering
  \caption{Parallel read time speedups over \textbf{single-threaded} fast\_matrix\_market for cold cache and warm cache experiments on SuiteSparse Matrix Collection matrices with at least one million entries.  Cold cache speedups are shown first, with warm cache speedups following in parantheses. ``P'' indicates parallel reading.}
  \begin{tabular}{l | r r | r r | r r}
    File Format & \multicolumn{2}{c|}{Avg Speedup} & \multicolumn{2}{c|}{Max Speedup} & \multicolumn{2}{c}{Min Speedup}\\
    \hline
    .mtx (P) & 2.3x & (3.3x) & 4.5x & (4.2x) & 0.9x & (0.6x)\\
    .coo.bsp (P) & 5.3x & (16.6x) & 24x & (54.7x) & 1.1x & (2.8x)\\
    .coo.bsp.gz (P) & 3.0x & (5.4x) & 28.6x & (26.6x) & 0.5x & (1.8x)\\
    .csr.bsp (P) & 6.1x & (19.7x) & 41.9x & (92.7x) & 1.2x & (3.3x)\\
    .csr.bsp.gz (P) & 3.2x & (5.9x) & 45.1x & (38.6x) & 0.6x & (1.6x)\\
  \end{tabular}
  \label{table:read_speedups_mt}
\end{table}

\begin{table}
  \centering
  \caption{Single-threaded warm cache read time speedups over splatt for FROSTT tensors with at least one million entries and fewer than 2 billion entries.}
  \begin{tabular}{l | r | r | r}
    File Format & Avg Speedup & Max Speedup & Min Speedup\\
    \hline
    .coo.bsp & 42.9x & 56.8x & 34.2x\\
    .coo.bsp.gz & 9.8x & 16.7x & 7x\\
    .csf.bsp & 56.3x & 190.9x & 34x\\
    .csf.bsp.gz & 9.6x & 16.6x & 5.7x\\
  \end{tabular}
  \label{table:read_speedups_tensor}
\end{table}

\subsection{Parallel Read Experiments}
Parallel file parsing often achieves faster performance than sequential parsing due to both the additional bandwidth available on modern SSDs, which often requires multiple threads to saturate, as well as the benefit of additional compute resources to handle computationally expensive parsing such as decompression or ASCII parsing with tools like \code{sprintf} and \code{iostream}.  fast\_matrix\_market is a multi-threaded capable parser, which can result in significant speedups over sequential parsing for computationally intensive Matrix Market parsing~\cite{Lugowski_fast_matrix_market_Fast_and_2023}.

Although there are proposals~\cite{elena_2024_13308713} to introduce multi-threaded support, the HDF5 library does not currently support multi-threading outside of mutex-based thread safety, which enforces sequential use of the HDF5 library and does not allow for efficient parallel reading.  This is true even for read-only use cases, as the HDF5 library has global state that is manipulated in various HDF5 calls.  Instead, we must use a \emph{multi-process} design for parallel reading.  To implement a parallel HDF5 reader, when reading an HDF5 dataset of size $n$, we fork off $p-1$ child processes to perform the read.  Each process then reads a block of approximately $n/p$ elements from the dataset, writing into a POSIX shared memory region.  Once all processes have completed their reads as determined using an atomic counter, the child processes exit and the parent process continues with execution.  This parallelization strategy uses multiple processes to read in each HDF5 dataset, allowing multiple processes to drive disk bandwidth higher as well as distribute computation-intensive parsing such as compressed reads.

Figure~\ref{fig:read_times_mt} shows the parallel read times for both warm and cold cache experiments.  As demonstrated by the plots, for parallel reading, Binsparse file formats again on average have much faster read times than Matrix Market files, particularly for large files.  For some small files, compressed Binsparse formats have similar or worse performance compared to Matrix Market.  This may be because of the relatively high overhead of forking off child processes and initializing the decompression library for files that take well below a second to parse.  However, for large files, we see significant speedups; for example, on the \code{Sybrandt/MOLIERE_2016} matrix, the largest matrix in the collection, reading the compressed CSR format is 1.9x faster (cold cache) than multi-threaded Matrix Market (7.5x speedup over sequential MTX), while the compressed CSR file is 2.7x smaller.

Similar to the single-threaded experiments, the uncompressed formats have faster parse times (4.7x faster than MTX for uncompressed CSR on \code{Sybrandt/MOLIERE_2016}) but at the cost of larger file sizes (only 2.2x smaller than MTX for uncompressed CSR).  Table~\ref{table:read_speedups_mt} shows the speedups achieved by both fast\_matrix\_market with multi-threading enabled and Binsparse's parallel reader over single-threaded fast\_matrix\_market.  As shown in the table, parallel reads of Binsparse formats are faster on average than multi-threaded fast\_matrix\_market for all formats, and in all cases for uncompressed formats.  Indeed, by comparing the speedup numbers in Tables~\ref{table:read_speedups} and ~\ref{table:read_speedups_mt}, we observe that single threaded reading of uncompressed Binsparse files achieves better performance than multi-threaded fast\_matrix\_market in all cases, as its minimum speedups are higher.

\begin{figure*}
  \centering
  \includegraphics[width=0.497\textwidth]{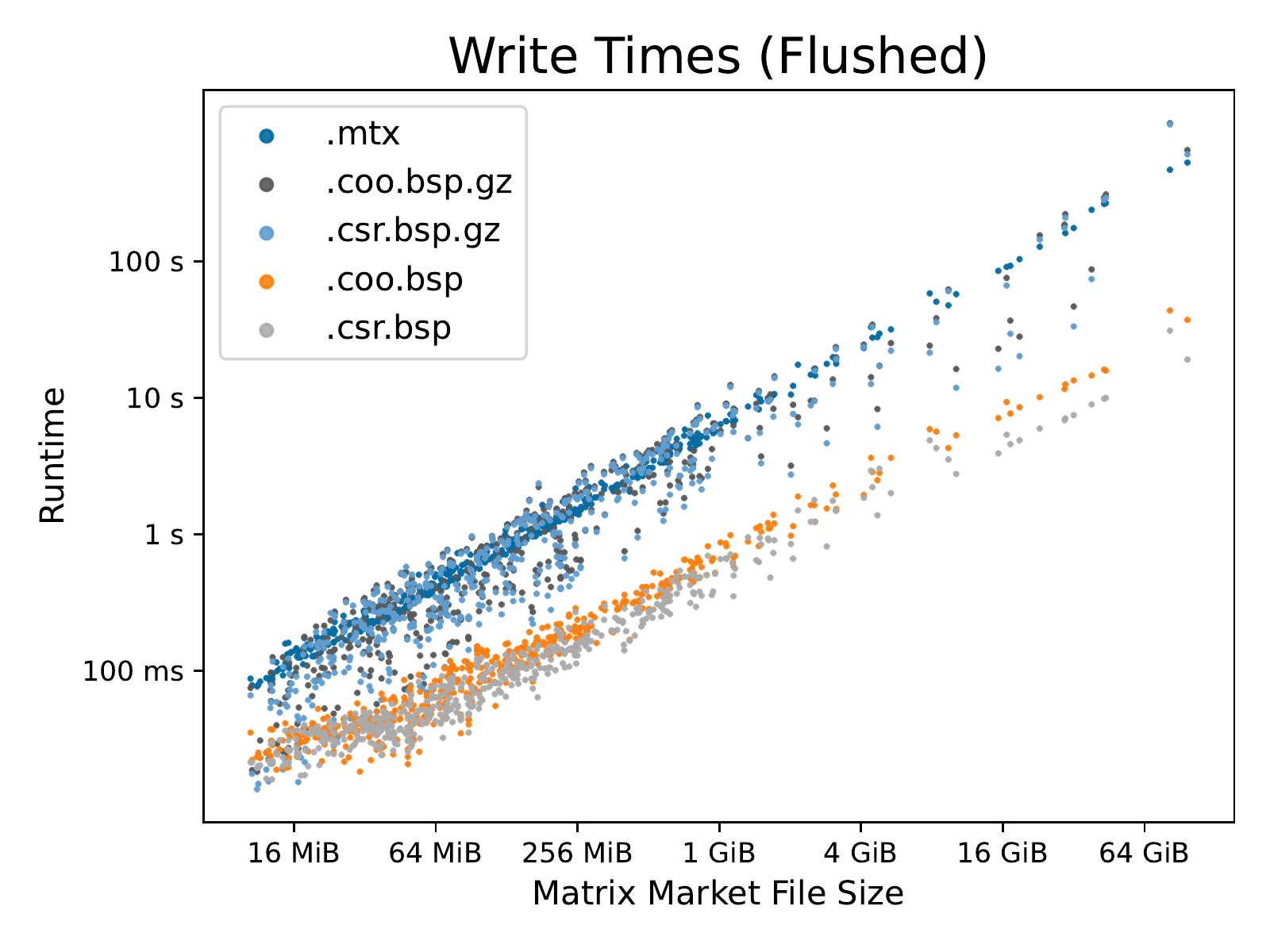}
  \includegraphics[width=0.497\textwidth]{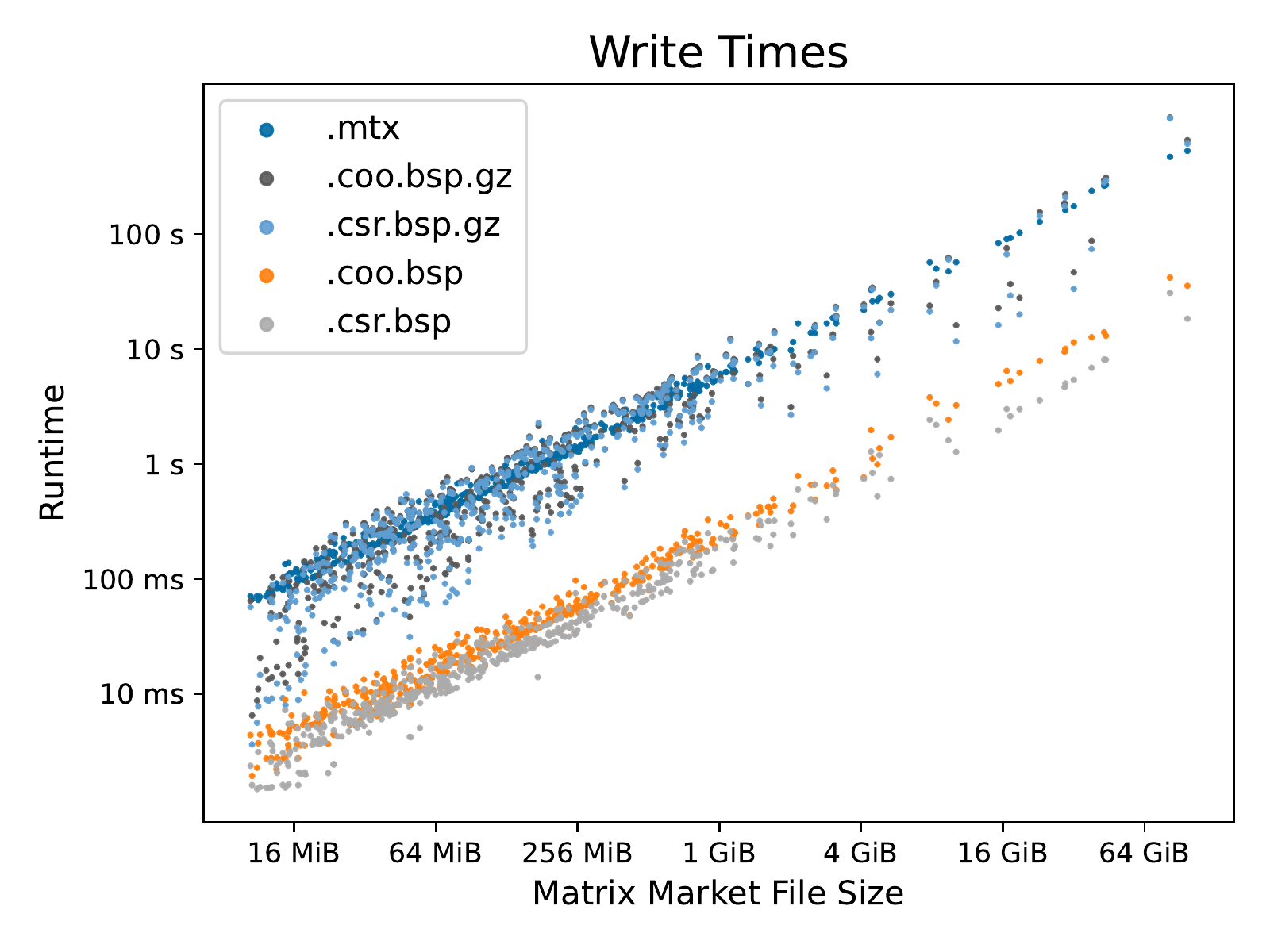}
  \caption{Single-threaded flushed and unflushed write times for all matrices in the SuiteSparse Matrix Collection with more than one million entries.  ``mtx'' indicates Matrix Market, while ``bsp'' indicates a Binsparse file format, ``csr'' indicates the CSR format is used, ``coo'' indicates the COO format is used, and ``gz'' indicates gzip compression is used.}
  \label{fig:write_times}
\end{figure*}

Comparing Binsparse's sequential and parallel read performance, performance is improved, on average, for cold reads of uncompressed Binsparse files, at around 1.4x improvement over sequential for both Binsparse CSR and COO.  However, warm cache reads of uncompressed formats on average achieve marginal speedups or even a slight slowdown compared to sequential Binsparse.  This is likely because of the overhead of starting multiple processes overwhelming an already very fast warm cache parsing speed.  As demonstrated by the maximum speedups, the speedup of parallel over sequential for large files is still significant, even for warm cache experiments.

The computationally expensive compressed formats achieve higher speedups on average for parallel parsing over sequential Binsparse for both warm cache and cold cache experiments, with around a 2.4x speedup for parallel over sequential parsing of compressed COO.  This is likely because the computationally expensive decompression process benefits more from the additional resources of multiple threads.

\subsection{Write Experiments}
Figure ~\ref{fig:write_times} plots time taken to write every sparse matrix in the SuiteSparse Matrix Collection with more than one million entries, including both flushed and unflushed experiments.  Each dot represents the time taken to write a single matrix to the filesystem, averaged over ten runs, with time in the y-axis and the size of the matrix's Matrix Market file on the x-axis.  As before, writing uncompressed Binsparse formats is faster than writing compressed formats, and in fact this effect is even more pronounced, since writing a compressed file is more computationally expensive than reading one~\cite{uber2022compression}.
This is demonstrated by the narrowing gap between flushed and unflushed experiments.  With compression dominating the runtime cost, the average speedups are 1.4x and 1.4x, respectively, for flushed and unflushed writes of the compressed Binsparse CSR format.  For some matrices, compressed writing is slower than Matrix Market writing, with up to a 2x slowdown.  However, for many matrices, compressed writing is still faster than Matrix Market, with up to a 18.3x speedup for compressed Binsparse CSR.  This is perhaps intuitive, as the compressed formats achieve very significant average file size reductions of 7.5x, with a maximum reduction of 53.6x for compressed Binsparse CSR.  Uncompressed Binsparse formats are still faster across the board, with even higher average speedups of 8.5x flushed and 31.2x unflushed for Binsparse CSR format.

\begin{table}
  \centering
  \caption{Single-threaded write time speedups over fast\_matrix\_market for flushed and unflushed experiments on SuiteSparse Matrix Collection matrices with at least one million entries.  Flushed speedups are shown first, with unflushed speedups following in parantheses.}
  \begin{tabular}{l | r r | r r | r r}
    File Format & \multicolumn{2}{c|}{Avg Speedup} & \multicolumn{2}{c|}{Max Speedup} & \multicolumn{2}{c}{Min Speedup}\\
    \hline
    .coo.bsp & 7x & (23.2x) & 16.9x & (41.4x) & 2.2x & (11.2x)\\
    .coo.bsp.gz & 1.3x & (1.3x) & 6x & (10.3x) & 0.5x & (0.5x)\\
    .csr.bsp & 8.5x & (31.2x) & 27.8x & (74.8x) & 2.9x & (15.2x)\\
    .csr.bsp.gz & 1.4x & (1.4x) & 8.9x & (18.3x) & 0.5x & (0.5x)\\
  \end{tabular}
  \label{table:write_speedups}
\end{table}

As discussed in Section~\ref{sec:minimization}, it should be noted that the Matrix Market files included in the SuiteSparse Matrix Collection have undergone an expensive minimization process, iterating through all possible printf specifier widths, to make them as compact as possible.  The fast\_matrix\_market parser, while very fast, does not perform this minimization process, and so the files it generates are larger.  For the matrices used in our experiments, the files written by fast\_matrix\_market using the default formatter were on average 2.2\% and up to 17\% larger than the reference Matrix Market files in the SuiteSparse Matrix Collection.

\section{Conclusions and Future Work}

In this paper we presented Binsparse, a specification for embeddable, cross-platform storage of sparse matrices and tensors to file.  We described the implementations of the specification using some existing binary file format toolchains and we examined how storing matrices with them results in significant file size reductions as well as much faster read and write times compared to ASCII-based formats.  The Binsparse specification provides a common framework and language for capitalizing on existing binary file format tools and ensures that sparse matrices and tensors stored with it can be used efficiently and widely.

One important area for future work is the implementation of a distributed Binsparse reader.  An implementation similar to CombBLAS's parser~\cite{azad2022combinatorial}, which has each process read an arbitrary subset of entries, followed by an all-to-all operation, would be both straightforward and general.  For formats that allow structured iteration, such as CSR, it should be possible to implement a distributed reader that only reads in a desired subset of elements, thus minimizing communication, as done in BCL's custom binary CSR format reader~\cite{brock2024}.  For distributed writing, it is likely that a blocked format that produces multiple Binsparse files will be desirable, due to the difficulty of writing to a single file using multiple processes.  These blocks could later be coalesced into a single Binsparse file.

\bibliographystyle{ACM-Reference-Format}
\bibliography{references}

\scriptsize
\noindent
\newline Optimization Notice: Software and workloads used in
performance tests may have been optimized for performance only on
Intel microprocessors.  Performance tests, such as SYSmark and
MobileMark, are measured using specific computer systems,
components, software, operations and functions.  Any change to any
of those factors may cause the results to vary.  You should
consult other information and performance tests to assist you in
fully evaluating your contemplated purchases, including the
performance of that product when combined with other products.
For more information go to \url{http://www.intel.com/performance}.

\noindent Intel, Xeon, and Intel Xeon Phi are trademarks of Intel Corporation in the U.S. and/or other countries.

\normalsize

\end{document}